\documentclass[12pt,preprint]{aastex}

\shorttitle{PDFs of molecular lines from turbulent clouds}

\shortauthors{BURKHART ET AL.}
\begin{document}

\title{The Effects of Radiative Transfer on the PDFs of Molecular MHD Turbulence}

\author{Blakesley Burkhart\altaffilmark{1}, V. Ossenkopf\altaffilmark{2},A. Lazarian\altaffilmark{1}, J. Stutzki\altaffilmark{2}}
\affil{$^1$ {Astronomy Department, University of Wisconsin, Madison, 475 N.  Charter St., WI 53711, USA}}
\affil{$^2$ {Physikalisches Institut der Universit\"{a}t zu K\"{o}ln, Z\"{u}lpicher Strasse 77, 50937 K\"{o}ln, Germany}}

\begin{abstract}
We study the effects of radiative transfer on the Probability Distribution Functions (PDFs) of simulations of magnetohydrodynamic turbulence in the widely studied  $^{13}$CO 2-1 transition. We find that the integrated intensity maps generally follow a log-normal distribution, with the cases that have $\tau \approx 1$ best 
matching the PDF of the  column density. 
We fit a  2D variance-sonic Mach number relationship to our logarithmic PDFs of the form  $\sigma_{ln(\Sigma/\Sigma_0)}^2=A\times ln(1+b^2{\cal M}_s^2)$ 
and find that, for parameter $b=1/3$, parameter $A$ depends on the radiative transfer environment.  
We also explore the variance, skewness, and kurtosis of the linear PDFs finding that higher moments reflect both higher sonic Mach number and lower optical depth. Finally, we  apply the Tsallis incremental PDF function and find 
that the fit parameters depend on both Mach numbers, but also are sensitive to the radiative transfer parameter space, with the 
$\tau \approx 1$ case best fitting the incremental PDF of the true column density.
We conclude that, for PDFs of low optical depth cases, part of the gas is always sub-thermally excited so that the spread of the line intensities exceeds the spread of the underlying column densities and hence the PDFs do not reflect the true column density.  Similarly, PDFs of optically thick cases  are dominated by the velocity dispersion and therefore do not represent the true column density PDF.  Thus, in the case of molecules like carbon monoxide, the 
dynamic range of intensities, structures observed and consequently, the observable 
PDFs,  are less determined by turbulence and more-often determined by radiative transfer effects.
\end{abstract}
 
 \section{Introduction}
\label{intro}

It is now accepted that the interstellar medium (ISM) of the Milky Way and other spiral galaxies is both magnetized and turbulent
and that this magnetohydrodynamic (MHD) turbulence
is one of the major physical processes that affect several fields of astrophysics
including star formation (see McKee \& Ostriker 2007 and references therein),
the structure formation and evolution of different phases of the interstellar medium 
(Vazquez-Semadeni 1994;  Ostriker et al. 2001; Elmegreen \& Scalo 2004;  Ballesteros-Paredes et al. 2007; Padoan et al. 2007; Chepurnov \& Lazarian 2010)
and magnetic reconnection (Lazarian \& Vishniac 1999; Eyink, Lazarian, \& Vishniac 2011).
The turbulent nature of ISM\footnote{The accepted paradigm
of MHD turbulence is based on the Goldreich \& Sridhar (1995) model which describes the properties of Alfv\'enic
turbulence. The decomposition in Alfv\'en, slow and fast modes was demonstrated in Cho \& Lazarian (2003), Cho et al. (2002) and Kowal \& Lazarian (2010).} 
gases 
is evident from a variety of observations including
electron density fluctuations, which demonstrate a Kolmogorov-type power law from kpc to AU \citep{arm95,chep10} and non-thermal broadening of emission and absorption lines, e.g. CO,
HI (Spitzer 1979; Stutzki \& Guesten 1990; Heyer \& Brunt 2004).

\begin{table}
\begin{center}

\begin{tabular}{ccccccccc}
\hline\hline
Run & $\approx {\cal M}_s$ & $\approx {\cal M}_A$ & $N$& $\tau$(norm)& $\tau (n$8250)& $\tau (n$9) &$\tau (x_{co}$-5)&$\tau (x_{co}$-8) \\
\tableline
1 &0.4&0.7&512&3.6&71&0.13&37&0.18\\
2&0.7&0.7&256&3.42&84&0.1&70&0.13\\
3&2&0.7&256&4.48&27&0.09&60&0.13\\
4&4&0.7&512&3&101&0.13&85&0.17\\
5&7&0.7&512&4.8&72&0.14&99&0.19\\
6&8&0.7&512&3.2&65&0.007&87&0.1\\
7&0.4&2&512&3.42&67&0.12&104&0.16\\
8&0.7&2&256&4.26&84.5&0.09&87.1&0.124\\
9&2&2&256&4.3&92.5&0.12&76.1&0.16\\
10&4&2&512&3.5&103&0.07&75&0.1\\
11&7&2&512&2.9&63&0.1&64&0.14\\
12&8&2&512&3.47&81&0.09&70&0.3\\
\hline\hline
\end{tabular}
\caption{Description of the twelve simulations used for this study, with model number given in column one. Columns two and three show the average sonic and Alfv\'enic Mach numbers. Column four shows the numerical resolution (N) of the simulations.  In columns five through nine we show the map-averaged line center optical depths for the different
values of our radiative transfer parameter space.  
For our radiative transfer parameter space we  vary both the number density scaling factor
(in units of $\rm{cm}^{-3}$, denoted with the symbol $n$) and the molecular abundance
($^{13}$CO/H$_2$, denoted with the symbol $x_{co}$) in order to change
excitation and optical depth $\tau$. To represent the
typical parameters of a molecular cloud we choose standard values for density,
$n=$275 $\rm{cm}^{-3}$, and abundance, $x_{co} = 1.5\times10^{-6}$ (what we will refer to as the normalized or norm parameter setup shown in column five).
We vary density and abundance  by values that are
factors of 30  larger and smaller as compared to the $\tau$(norm) values.
Thus column six, denoted by $n8250$, has parameters: $n=$8250 $\rm{cm}^{-3}$, and abundance, $x_{co} = 1.5\times10^{-6}$.
Column seven, denoted by $n9$, has parameters: $n=$9 $\rm{cm}^{-3}$, and abundance, $x_{co} = 1.5\times10^{-6}$.
Column eight, denote by $x_{co}-5$, has parameters: $n=$275 $\rm{cm}^{-3}$, and abundance, $x_{co} = 4.5\times10^{-5}$.
Column nine, denoted by $x_{co}-8$, has parameters: $n=$275 $\rm{cm}^{-3}$, and abundance, $x_{co} = 5\times10^{-8}$. 
\label{fig:radtab}}
\end{center}
\end{table}

The molecular medium, including giant molecular clouds (GMCs), represents the densest part of the turbulent cascade.  
Many authors have shown that molecular clouds are  in fact dominated by magnetized turbulent motions, as is evident from line-width size relations,
the power spectrum and the fractal nature of the observed morphology (Larson 1981; Stutzki et al. 1998).
 However, the exact role of turbulence in GMCs regarding the process of stellar birth is under current debate.  
Turbulence is a vital component in several theoretical scenarios of the star formation
process  such as 
star formation resulting from density compressions in supersonic turbulence
(Elmegreen 1993; Federrath \& Klessen 2012) and local core collapse in globally turbulent pressure supported clouds  (Jeans 1902; Hunter 1979;  Klessen et al. 2000).
 A recently proposed model of magnetic flux transport out of clouds, namely, the model of ``turbulent reconnection diffusion" (Lazarian 2005; 
Lazarian et al. 2010; Santos-Lima et al. 2010, 2011) is based on the theory of fast reconnection of magnetic fields in a  {\it turbulent} medium 
(Lazarian \& Vishniac 1999).

Despite the recognition that turbulence (particularly supersonic turbulence)
and magnetic fields are important to studies of molecular clouds and star formation, the community still faces substantial challenges in studying 
and characterizing the MHD properties in a quantitative way. 
This is, in part, because the fundamental parameters of interstellar turbulence,
such as the sonic  and Alfv\'enic Mach numbers, as well as the virial parameter and turbulence driving mechanisms (i.e. the forcing parameters), are difficult to determine due to the challenges of imaging small scale turbulence (Bertoldi \& McKee 1992; Federrath et al. 2010; Gaensler et al. 2011; Burkhart et al. 2012a) and because measurements of magnetic fields are are still marginal and time consuming.

\begin{figure*}[tbh]
\centering
\includegraphics[scale=.33]{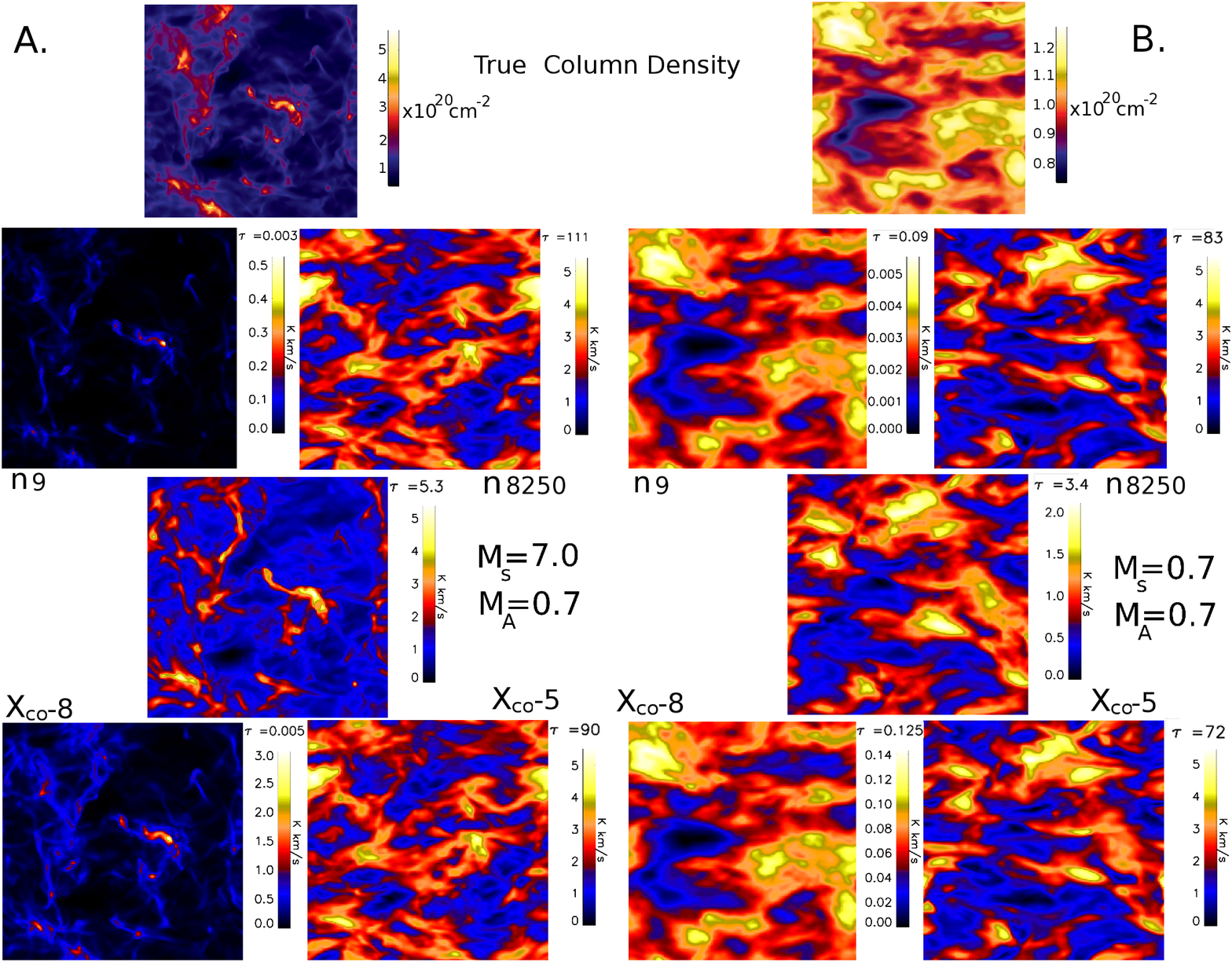}
\caption{ $^{13}$CO 2-1 Integrated intensities for a range of optical depths, abundances ($x_{co}$), and initial densities ($n$) for: panel A (left), a supersonic sub-Alfv\'enic model (model number 5 from Table~\ref{fig:radtab})
and panel B, a subsonic sub-Alfv\'enic model (model number 2 from Table~\ref{fig:radtab}).  The column density maps for each case are shown in a single panel on the very top row.
For each simulation, the middle single panels shows the standard ($\tau$ norm) case, the two bottom panels are for varying the abundances 30 times smaller ($x_{co}-8$, left) and 30 times larger ($x_{co}-5$, right),  and the second to top
two panels show varying the density 30 times smaller ($n 9$, left) and 30 times larger ($n 8250$, right).}
\label{fig:supercods}
\end{figure*}

In addition to observational data, the analysis of simulations have proven to be very useful for studying turbulence in the diffuse ISM and in star forming regions.
The most common statistical tool used for studies of turbulence is the spatial power spectrum of column density maps  or electron scintillation and scattering measurements (Armstrong et al. 1995; Elmegreen \& Scalo 2004)
although more recently several studies have undertaken measurements of the velocity power spectrum (Lazarian 2009). 
More recently, the power spectrum of column density maps has also been used to estimate the star formation efficiency rates 
in numerical simulations (Federrath \& Klessen 2013).
Although the power spectrum is useful for obtaining information about energy transfer over scales, it does not
provide a full picture of turbulence, partially because it
only contains information on Fourier amplitudes. In light of this, many other techniques have been developed to study and parametrize observational magnetic
turbulence. These include higher order spectra, such as
the bispectrum, higher order statistical moments, topological techniques (such as genus), clump and hierarchical structure algorithms (such as dendrograms), principle component analysis, Tsallis function studies for ISM turbulence, and structure/correlation functions
as tests of intermittency and anisotropy (for examples of
such studies see Heyer \& Schloerb 1997;  Chepurnov et al. 2008; Burkhart et al.
2009; Chepurnov \& Lazarian 2009; Kowal, Lazarian, \&
Beresnyak 2007; Goodman et al. 2009a; Esquivel \& Lazarian 2010; Roman-Duval et al. 2011; Tofflemire, Burkhart, \& Lazarian 2011;  Burkhart et al.
2012b).  Wavelets methods, and variations on them such
as the $\Delta$-variance method, have also been shown to be
very useful in characterizing the distribution of inhomogeneities in data (see
Ossenkopf \& McLow 2002; Ossenkopf, Krips, \& Stutzki 2008a, 2008b; Schneider et al. 2011).

However, many of the statistical studies being conducted fail to take into account key radiative transfer properties and observational
constraints, such as the varying optical depth,  when investigating
the structure and statistics of turbulence.
This is particularly true for star forming molecular clouds, as 
absorption effects are important for all abundant molecular tracers, like e.g. $^{12}$CO and $^{13}$CO.
Furthermore, recent studies have
brought into question whether the emission of CO or its isotopes can
provide a good representation of the true density distribution
in molecular clouds and under which conditions do they 
really trace H$_2$ (see Goodman et al. 2009b; Glover et al. 2010; Shetty et al. 2011a,b).
Four different effects may change the so-called $X$-factor, i.e. the ratio between the
CO intensity and the column density of molecular ($H_2$) gas. These are variations
of the abundance of CO relative to H$_2$, the temperature of the molecular
cloud, the velocity range, and density effects: 
subcritical collisional excitation in thin gas, and saturation due
to high optical thickness for large column densities. Previous systematic
studies \citep{shetty11a, shetty11b} combined all effects but focused on
the impact of a changing molecular abundance. To obtain a more complete
description in terms of the impact of the different physical effects on the
measured structure of a molecular cloud, we investigate here in detail
the impact of sub-thermal excitation and radiative transfer effects on the
emission of turbulent molecular clouds. We use isothermal MHD simulations
to simulate molecular line maps of the $^{13}$CO 2-1 transition from the
models comparable to the observed maps. We generate the molecular line maps
using the Ossenkopf (2002) radiative transfer algorithm (SimLine-3D). 
We then investigate the resulting CO line mps, in particular their 
Probability Distribution Functions (PDFs) 
and their descriptors, such as variance, skewness, and kurtosis and
the Tsallis function fit, and study how these quantities 
change when changing the optical depth, density, and abundances (see Table \ref{fig:radtab} for details on our parameter space).

For this paper, we concern ourselves only with the 2D quantities (column density, integrated intensity) that are available from direct observations.
We will denote the 2D column density or integrated intensity distribution as $\Sigma$.  We will use $\zeta$ to stand for the logarithm of the 2D quantities, (i.e. $\zeta=ln(\Sigma/\Sigma_0)$, 
where $\Sigma_0$ is the mean value of  $\Sigma$).

The paper is organized as follows. In \S~\ref{sec:pdf} we describe the PDF methods used in this paper.
In \S~\ref{sec:Sims} we describe the simulations and outline the main points of the radiative transfer algorithm.
In \S~\ref{sec:pdfs} we discuss the use of PDFs on turbulent clouds and explore the PDFs of our models.
In \S~\ref{sec:lin} we discuss the variance vs. sonic Mach number relationship for our models.
In \S~\ref{sec:skew} we discuss the relationship between the skewness and kurtosis and the sonic Mach number for our models.
In \S~\ref{sec:tsal} we apply the incremental PDF functional fit known as the Tsallis distribution to our models and investigate
the relationship the fit parameters have with the optical depth, sonic and Alfv\'enic Mach numbers.
Finally, in  \S~\ref{sec:dis} we discuss our results followed by the conclusions in \S~\ref{sec:con}.

\section{Probability Distribution Functions as a Tool for Turbulence Studies}
\label{sec:pdf}
PDFs are one of the statistical tools developed to study turbulence in both diffuse and self-gravitating clouds. 
PDFs and their quantitative descriptors
have been used to study turbulence in a variety of astrophysical contexts including the diffuse ISM
(Hill et al. 2008; Berkhuijsen \& Fletcher 2008), the solar wind (Burlaga et al. 2004a) and the molecular ISM (Padoan et al. 1997; Kainulainen \& Tan 2012).

Several authors have suggested that PDFs of column density data can be used to predict important parameters of ISM turbulence in the diffuse and molecular medium
such as the sonic Mach number (Burkhart et al. 2009, 2010; Price, Federrath, \& Brunt 2011), the star formation rate in GMCs (Krumholz \& McKee 2005; Federrath \& Klessen 2012), and the initial
mass function (Hennebelle \& Chabrier 2008, 2011).
Two predominant ways of exploring the PDFs of turbulence have been investigated: one focuses on the statistics of linear column density while the other
takes advantage of the generally log-normal nature of turbulent fields. In either case, researchers have either calculated the moments of the PDF distribution
such as the variance, skewness, and/or kurtosis or fitted a function to the PDF and used the resulting fit parameters as descriptors of turbulence
(e.g. the Tsallis function, Esquivel \& Lazarian 2010; Tofflemire, Burkhart, \& Lazarian 2011).   

In regards to the moments, several studies have numerically investigated the relationship between the \textit{3D density}
PDF width (i.e. the variance or standard deviation) and the root mean square (rms) sonic Mach number for linear and natural-log density (Padoan et al. 1997; Price, Federrath, \& Brunt 2011; Molina et al. 2012).  
For a distribution $X={X_i, i=1...N}$ the mean value and variance are defined as:
$\overline{X}=\frac{1}{N}\sum_{i=1}^N {\left( X_{i}\right)}$
and $\sigma_{X}^2= \frac{1}{N} \sum_{i=1}^N {\left( X_{i} - \overline{X}\right)}^2$.
The primary advantage of taking the logarithm of the density field is that turbulent density fields are generally log-normal, and thus researchers can take advantage of analytically relationships of log-normal distributions.
A key assumption in these models is that there is a linear  relationship between the variance in linear density and the sonic Mach number, i.e,
\begin{equation}
 \sigma_{\rho/\rho_0}^2=b^2{\cal M}_s^2
\label{eq:linvar}
\end{equation}
where $\rho_0$ is the mean value of the 3D density field, $b$ is a constant of order unity and $\sigma_{\rho/\rho_0}$ is the standard deviation of the density field
normalized by its mean value.  When taking the logarithm of the normalized density field this relationship becomes:
\begin{equation}
 \sigma_{s}^2=ln(1+b^2{\cal M}_s^2)
\label{eq:logvar}
\end{equation}
where $s=ln(\rho/\rho_0)$ and $\sigma_{s}$ is the standard deviation of the logarithm of density (not to be confused with $\sigma_{\rho/\rho_0}$).
Reported values for $b$ vary widely for different numerical studies and under different turbulent driving schemes.  Values of $b$ from numerical studies are found to be between 0.26 and unity.
Generally, the value depends on the driving of the turbulence in question with $b=1/3$ for solenodial forcing and $b=1$ for compressive driving
(Nordlund \& Padoan 1999; Federrath, Klessen, \& Schmidt 2008).  

A key limitation of many of the variance-Mach number studies mentioned above is that the relationships
are derived only for 3D density, which is not observable.  Recently, several authors have attempted to solve this problem.
A study by Brunt, Federrath, \& Price (2010) devised a method to reconstruct the 3D variance
density from the projected 2D variance, however the study found a $\approx40\%$ error for different sight-lines for the case of ideal MHD turbulence and did not include
effects of self-gravity or radiative transfer (which can significantly alter the distribution of integrated intensity).   
A study by Burkhart \& Lazarian (2012) modified Equation \ref{eq:logvar} to be \textit{directly} applied to 2D column density maps.

The variance is not the only PDF descriptor that can be related to the sonic Mach number.  Studies of the higher order moments of density and column
density (for linear distributions) by Kowel et al. (2007) and Burkhart et al. (2009, 2010) showed that the skewness and kurtosis have a nearly linear relationship with the sonic Mach number.
The 3rd  order moment or skewness is defined as:
\begin{equation}
\gamma_{X} = \frac{1}{N} \sum_{i=1}^N{ \left( \frac{X_{i} - \overline{X}}{\sigma_{X}} \right)}^3
\label{eq:skew}
\end{equation}
and the 4th order moment kurtosis as:
\begin{equation}
 \beta_{X}=\frac{1}{N}\sum_{i=1}^N \left(\frac{X_{i}-\overline{X}}{\sigma_{X}}\right)^{4}-3
\label{eq:kurt}
\end{equation}
Skewness describes the asymmetry of a
distribution about its mean.  The
fourth order moment, kurtosis, is a measure of a distribution's sharpness or flatness
compared to a Gaussian distribution.  Skewness and kurtosis are therefore unitless numbers which require
no normalization between simulations/observations (variance, however, is scale dependent).  They also require no assumption of log-normality 
for the PDF in question.
As the sonic Mach number increases, shocks create density enhancements  and hence the skewness and kurtosis
increase.  This  reflects a departure from Gaussian subsonic distributions, which have skewness/kurtosis $\approx$ 0. 
 
 Additional PDF descriptors used for studies of ISM turbulence include the Tsallis function. The Tsallis distribution is a function
that can be fit to incremental PDFs of turbulent density,
magnetic field, and velocity. One of the Tsallis fit's first practical application for studies of turbulence were on PDFs of the solar wind.  Burlaga
and collaborators applied the Tsallis distribution to the solar wind magnetic and velocity field data (Burlaga \& Vi\~{n}as 2004a, 2004b, 2005a, 2005b, 2006) to describe the temporal variation
in PDFs  measured by the
\emph{Voyager 1} \& \emph{2} spacecrafts. Esquivel \& Lazarian (2010) and Tofflemire, Burkhart, \& Lazarian (2011)
 used the Tsallis statistics to describe the spatial variation in PDFs of density, velocity, and magnetic field of
 MHD simulations similar to those used here. Both efforts
found that the Tsallis distribution provided adequate fits to their PDFs
and gave insight into the sonic and Alfv\'enic Mach numbers of MHD turbulence.
 
\begin{figure}[tbh]
\centering
\includegraphics[scale=.7]{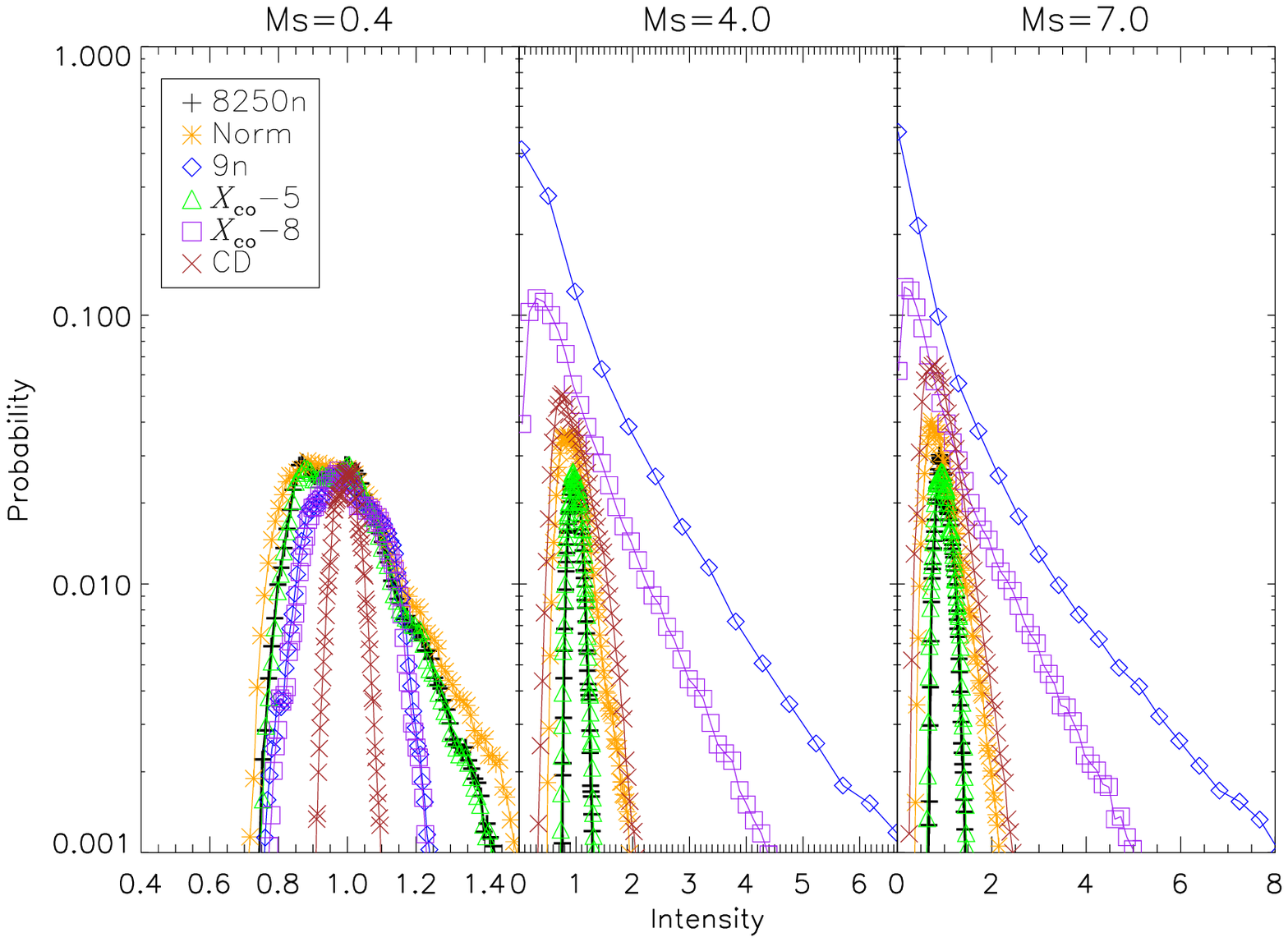}
\caption{ PDFs of $\Sigma/\Sigma_0$.  The far left panel has ${\cal M}_s\approx0.4$ the central panel has ${\cal M}_s\approx4.0$, and the far right panel
has ${\cal M}_s\approx7.0$.  All models have ${\cal M}_A=2.0$.  }
\label{fig:norm_pdf}
\end{figure}
To our knowledge, no numerical study has been performed to address the effects of radiative transfer and optical depth on the above mentioned 
PDF descriptors and their relations to the sonic Mach number.

\section{MHD Simulations and Radiative Transfer}
\label {sec:Sims}

We generate  3D numerical simulations of isothermal compressible (MHD)
turbulence by using the the Cho \& Lazarian  (2003) MHD code and varying the input
values for the sonic and Alfv\'enic Mach number.

 We briefly outline the major points of their numerical setup (for more details see Cho et al. 2002, Cho \& Lazarian 2003).

The code is  a second-order-accurate hybrid essentially
nonoscillatory (ENO) scheme (Cho \& Lazarian 2003) which solves
the ideal MHD equations in a periodic box:
\begin{eqnarray}
 \frac{\partial \rho}{\partial t} + \nabla \cdot (\rho {\bf v}) = 0, \\
 \frac{\partial \rho {\bf v}}{\partial t} + \nabla \cdot \left[ \rho {\bf v} {\bf v} + \left( p + \frac{B^2}{8 \pi} \right) {\bf I} - \frac{1}{4 \pi}{\bf B}{\bf B} \right] = {\bf f},  \\
 \frac{\partial {\bf B}}{\partial t} - \nabla \times ({\bf v} \times{\bf B}) = 0,
\end{eqnarray}
with zero-divergence condition $\nabla \cdot {\bf B} = 0$,
and an isothermal equation of state $p = C_s^2 \rho$, where
$p$ is the gas pressure.
On the right-hand side, the source term $\bf{f}$ is a random
large-scale driving force (${\bf f}= \rho d{\bf v}/dt$).  
We drive turbulence solenoidally, at wave scale $k$ equal to about 2.5
(2.5 times smaller than the size of the box). This scale defines
the injection scale in our models in Fourier space to minimize
the influence of the forcing on the generation of density structures.
Density fluctuations are then generated by the interaction of MHD waves.
 The time $t$ is in units of the large eddy turnover time
($\sim L/\delta V$) and the length in units of $L$, the scale
of energy injection.
The magnetic field consists of the uniform background field and a
fluctuating field: ${\bf B}= {\bf B}_\mathrm{ext} + {\bf b}$. Initially ${\bf b}=0$.

We divided our models into two groups corresponding to
sub-Alfv\'enic ($B_\mathrm{ext}=1.0$) and
super-Alfv\'enic ($B_\mathrm{ext}=0.1$) turbulence.
For each group we compute several models with different values of
gas pressure, which is our control parameter that sets the sonic Mach number (see Table\ref{fig:radtab}, second column).
We run compressible MHD turbulent models, with 256$^3$ and 512$^3$ resolution,
for $t \sim 5$ crossing times, to guarantee full development of the energy cascade.
Since the saturation level is similar for all models and we solve the isothermal MHD
equations, the sonic Mach number is fully determined by the value of the isothermal
sound speed, which is our control parameter. 
The models are listed and described in Table \ref{fig:radtab}.

After we generate the simulations, including the full 3D density and velocity
cubes, we apply the SimLine-3D radiative transfer algorithm (Ossenkopf 2002) for
the $^{13}$CO 2-1 transition. The code computes the local excitation of
molecules from collisions with the surrounding gas and from the radiative
excitation at the frequencies of the molecular transitions through line and
continuum radiation from the environment. Instead of a $\Lambda$-iteration
solving the mutual dependence of the radiative excitation of every point in
a cloud on the excitation of every other point in the cloud, the code uses
two approximations to describe the radiative interaction. First it computes
a local radiative interaction volume limited by the velocity gradients (LVG). Cells
with line-of-sight velocities different by more than the thermal line width
cannot contribute to the excitation of the considered molecule, i.e. the
interaction length is limited by the ratio of the thermal line width to the
velocity gradient. This is a particularly useful approximation for supersonic
turbulence. For the interaction of more remote points that accidentally
have the same line of sight-velocity, the second approximation applies, using
the average isotropic radiation field in the cube for the excitation at
every individual point. This best covers all cases with isotropic
structure, i.e. all turbulence simulations with good statistics not
dominated by few large-scale structures.
For all our cases, the approximations should be very good, so that we expect
an accuracy of better than 10\,\%. This level of accuracy is sufficient for
comparison with observational data as drifts in the receiver system and the atmosphere
and the resulting temporal variation of the calibration parameters typically also
provide calibration errors of that magnitude.
\begin{figure}[tbh]
\centering
\includegraphics[scale=.6]{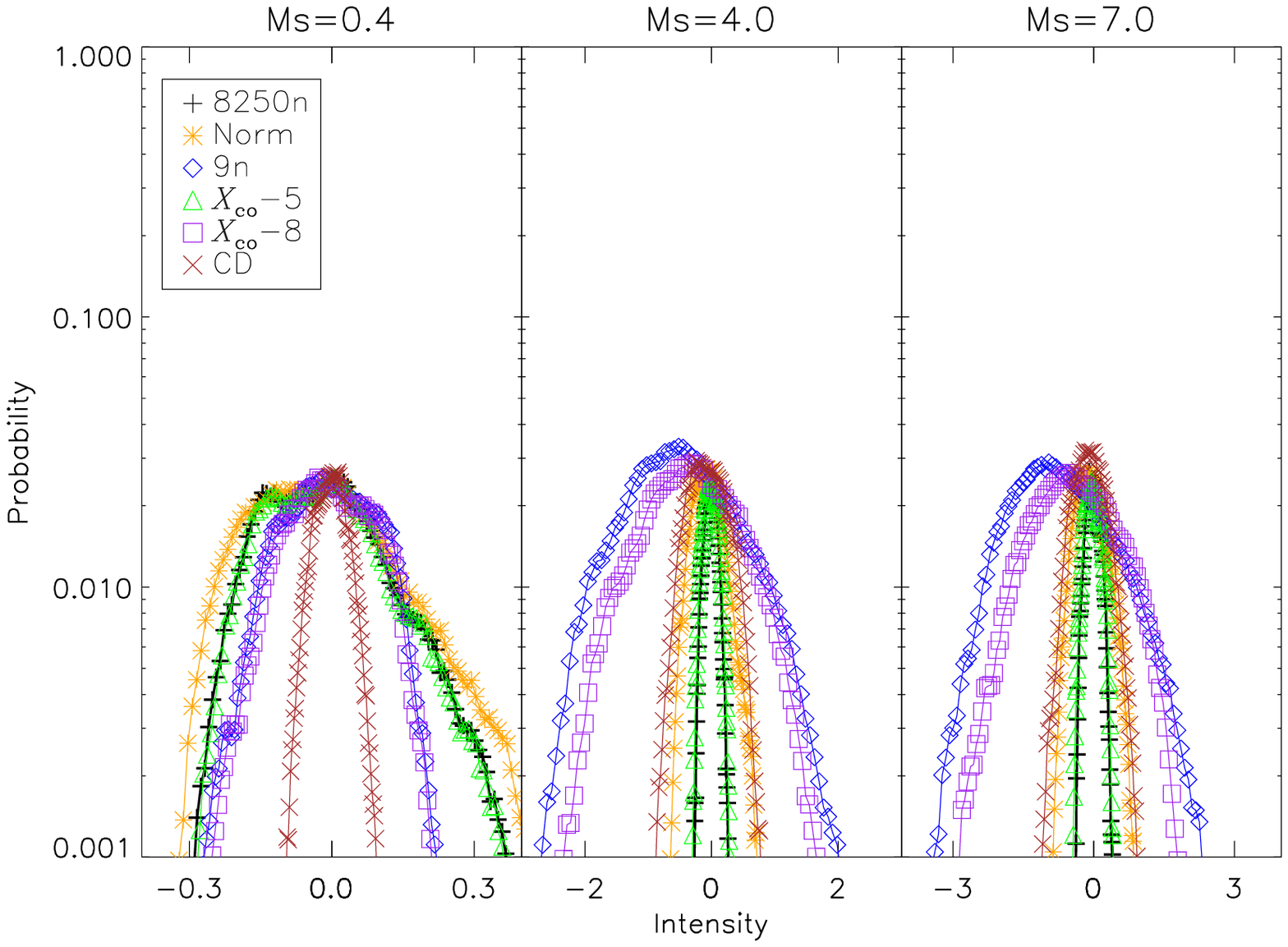}
\caption{PDFs of $\zeta$=ln($\Sigma/\Sigma_0$). The far left panel has ${\cal M}_s \approx 0.4$ the central panel has ${\cal M}_s \approx 4.0$, and the far right panel
has ${\cal M}_s \approx7.0$.  All models have ${\cal M}_A \approx 2.0$. }
\label{fig:lognorm_pdf}
\end{figure}

For our tests we choose a total cube size of 5pc and a gas temperature of 10K.
The cube is observed at a distance of 450pc with a beam FWHM of 18'' and a
velocity resolution of $0.05\rm{kms}^{-1}$.  We note that the original simulations set the sound speed by
varying the isothermal pressure and holding the RMS velocity constant. However, now we set a constant temperature of 10K for all the simulations.
In this case, it makes sense to hold the sound speed constant and vary the RMS velocity in order to set the isothermal sonic Mach number.  However, as we
show in the appendix, the resulting PDFs are unaffected by the rescaling of the velocity field, and the parameters of critical importance remain the optical depth and sonic Mach number.  Additionally, the radiative transfer has a dependency on the temperature, however here we concern ourselves with the density regime of GMCs that is known to be approximately isothermal, i.e. at intermediate densities, such as the  norm case in Table 1, which are larger than the transition density to the CNM but smaller than the densities leading to star formation.

We vary both the number density scaling factor
(in units of $\rm{cm}^{-3}$, denoted with the symbol $n$) and the molecular abundance
($^{13}$CO/H$_2$, denoted with the symbol $x_{co}$) in order to change
excitation and optical depth $\tau$. To represent the
typical parameters of a molecular cloud we choose standard values for density,
$n=$275 $\rm{cm}^{-3}$, and abundance, $x_{co} = 1.5\times10^{-6}$  
(this is what we will refer to as the normalized or norm parameter setup), 
providing a slightly optically thick $^{13}$CO 2-1 line at the highest densities with line
center optical depths of a few ($\tau$(norm) in Table \ref{fig:radtab} in
column 5). To investigate the impact of line saturation and subthermal
excitation we vary density ($n$) and abundance ($x_{co}$) by
factors of 30 to larger and smaller values compared to the $\tau$(norm) values.
We list all the parameters of the complete model set in Table~\ref{fig:radtab}.

We show examples of our synthetic  $^{13}$CO 2-1 integrated intensity maps for a range of optical depths, abundances ($x_{co}$), and initial
 densities ($n$) for a supersonic case (model 6) and a subsonic case (model 2) in Figure  \ref{fig:supercods} .  Both cases are sub-Alfv\'enic. 

Increasing density or abundance (far right images in panels A and B of Figure \ref{fig:supercods}) leads to similar effects of optically very thick lines with a peak optical depth around $\tau\approx 100$.
The map is dominated by the velocity dispersion, not the true column density structure. For lowered abundances (bottom left  images in panels A and B of Figure \ref{fig:supercods}), we get a more accurate
representation of the true column density structure. For lowered densities (top left images in panels A and B of Figure \ref{fig:supercods}), the regions of low density become 
invisible in $^{13}$CO because the molecule is no longer thermally excited.

Examination of Figure \ref{fig:supercods} clearly show that  different sonic Mach numbers provide
different distributions of the integrated intensity maps for $^{13}$CO 2-1 emission. 
 However, comparison between different ranges of average density/abundance ratio 
(and as a result, different values of optical depth) also show differences in the distribution and 
intensity values of the integrated intensity maps (and therefore the resulting column density maps of $H_2$ for a given X factor).  
In fact, in many cases presented in Figure \ref{fig:supercods}  (especially the highly optically thick cases), it is clear that the radiative transfer 
dominates over turbulence in terms of the distribution and structures present. 
Thus for carbon monoxide, the common  statistics of turbulence studies must be revisited to include 
these effects.   In the next sections we will investigate the effects of optical depth on the PDF moments 
and incremental PDF Tsallis fits  in order to determine 
over what ranges of optical depth these statistics are still 
sensitive to the Mach numbers for $^{13}$CO 2-1 emission.   

\section{PDFs of $^{13}\rm{CO}$ Turbulent Maps}
\label{sec:pdfs}
We calculate the PDFs for all models listed in Table \ref{fig:radtab} for both the linear distribution and taking the natural logarithm of the maps.
We make plots of the PDFs of the models for super-Alfv\'enic turbulence.  We plot the linear PDFs (i.e. PDFs of $\Sigma/\Sigma_0$) of the column density (red lines) and integrated
intensity in Figure \ref{fig:norm_pdf}.  The panels are ordered subsonic to supersonic going left to right. 
Different colored lines and symbols represent our varying radiative transfer parameter space.
For the subsonic case (far left panel) it is clear that \textit{none} of the models that include radiative transfer
exactly match with the subsonic column density PDF of $\Sigma/\Sigma_0$, which is due to the fact that mutual radiative pumping takes over the role of thermalization of the lines.
As the sonic Mach number increases (center panel with ${\cal M}_s\approx4.0$ and far right panel with ${\cal M}_s\approx7.0$)
we can also see a general departure from the column density PDF as the radiative transfer environment alters the integrated intensities.  Additionally, it is clear
that the 'norm' (yellow asterisk) model PDF with $\tau\approx1$ fits very closely to the column density PDF.

The column density PDF becomes increasingly skewed and kurtotic as the sonic Mach number increases
compared to the Gaussian-like distribution seen in the subsonic panel.  This is similar to findings from past works such as
Kowal et al. (2007) and Burkhart et al. (2009, 2010), who suggested that these trends are due to high density regions created by shocks which skew
and peak the distribution away towards higher density values.  However, neither of these 
studies explored distributions of integrated intensity maps.
As the optical depth decreases, either due to decreasing the abundance (purple squares) or the average density (blue diamonds), the distribution becomes increasingly
skewed and kurtotic compared with the column density.  This is because the emission is concentrated only in the regions of higher density and abundance
while the thin gas is not strongly emitting, creating low emission holes. 
For cases of higher optical depth, either from increasing the density (black plus signs) or increasing the abundance (green triangles)
the distributions are significantly less skewed and peaked compared with the column density.  This is due to the effects of self-absorption in the gas. 
The cloud becomes optically thick before it can be fully sampled, thus masking out any shock enhancements that could be
seen in the cloud. Thus in the case of high optical depths, the distributions are similar to the low Mach-number cases.

\begin{figure}[tbhp]
\centering
\includegraphics[scale=.8]{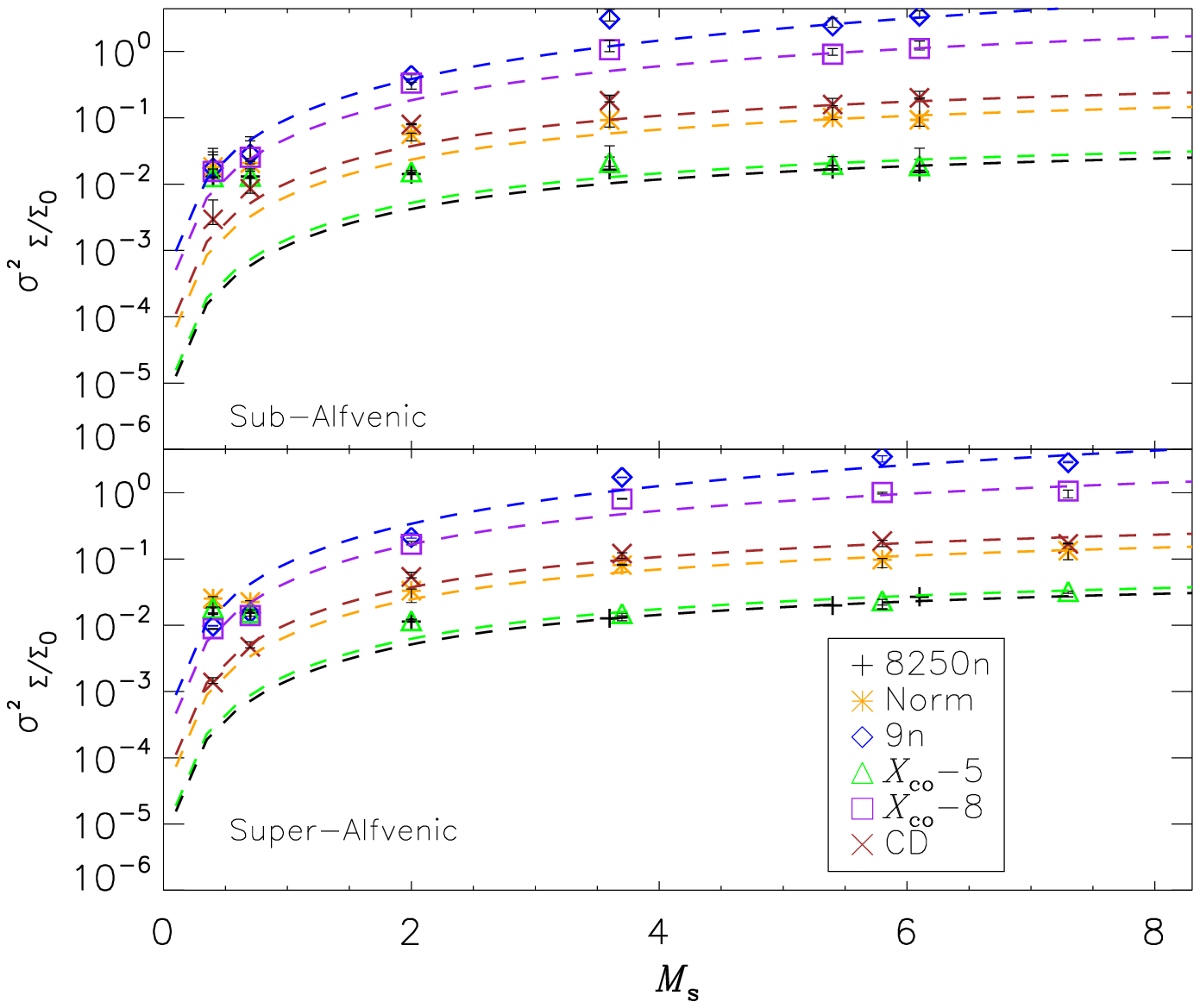}
\caption{Linear variance ($\sigma^2_{\Sigma/\Sigma_0}$) vs. RMS ${\cal M}_s$ for all our simulations with the best fit given by Equation \ref{eq:linvar_cd} (dashed lines). Sub-Alfv\'enic
cases are shown in the top panel, while super-Alfv\'enic cases are presented in the bottom panel. The true column density is shown with the red line.
While Equation \ref{eq:linvar_cd} fits fairly well for supersonic simulations, values of $A$ are clearly dependent on the radiative transfer environment, as shown in Table 2.  }
\label{fig:var_lin}
\end{figure}

We make plots of the  PDFs for the logarithm of the distribution ($\zeta=ln(\Sigma/\Sigma_0)$)
of the models for super-Alfv\'enic turbulence in the top plot of Figure \ref{fig:lognorm_pdf}. 
We plot the  PDFs of the logarithm of column density (red lines) and logarithm of
integrated intensity with the same models as in Figure \ref{fig:norm_pdf}.  

All distributions appear approximately log-normal, however some deviations
can be seen in the subsonic PDFs (far left panel).  
With line saturation, the PDF can be expected to be asymmetric, with a tendency to pile up at larger intensities.
This effect was also observed in Shetty et al. 2010.
For the subsonic case, all models with radiative transfer
show slightly larger widths than the true column density, regardless of optical depth. 

In these supersonic cases (which all generally appear log-normal), the norm case with a mixed optically thin/thick
medium  (orange stars) has a width that is generally the closest to the column density.  The two optically thin cases (blue diamonds and purple
squares) have very large widths and begin to deviate from log-normal.  This is again due to quenching of the emission due to a limited range of densities/ abundances, which in turn increases the variance in the distribution
due to larger contrast between the few areas with enough density/tracer material and the invisible gas.  The optically thick cases (green triangles and orange stars) have smaller
PDF widths than the true column density.

Figures  \ref{fig:norm_pdf} and \ref{fig:lognorm_pdf} show that the PDFs of the 2D distribution of MHD turbulence as seen from
observations will not only depend on the sonic Mach number, but also on the radiative transfer environment and the optical depth
of the medium.  This fact has not been accounted for in previous studies, and may affect the derived relationships
between the sonic Mach number and the variance, skewness, and kurtosis of linear and logarithmic column density/integrated intensity.
It may also explain the discrepancy and uncertainties
between the PDFs of numerical simulations to those of observations (see Goodman et al. 2009b; Price, Federrath \& Brunt 2011;
Burkhart et al. 2009,2010;  Brunt 2010).  We address these issues in the next two sections.

\begin{figure}[tbh]
\centering
\includegraphics[scale=.78]{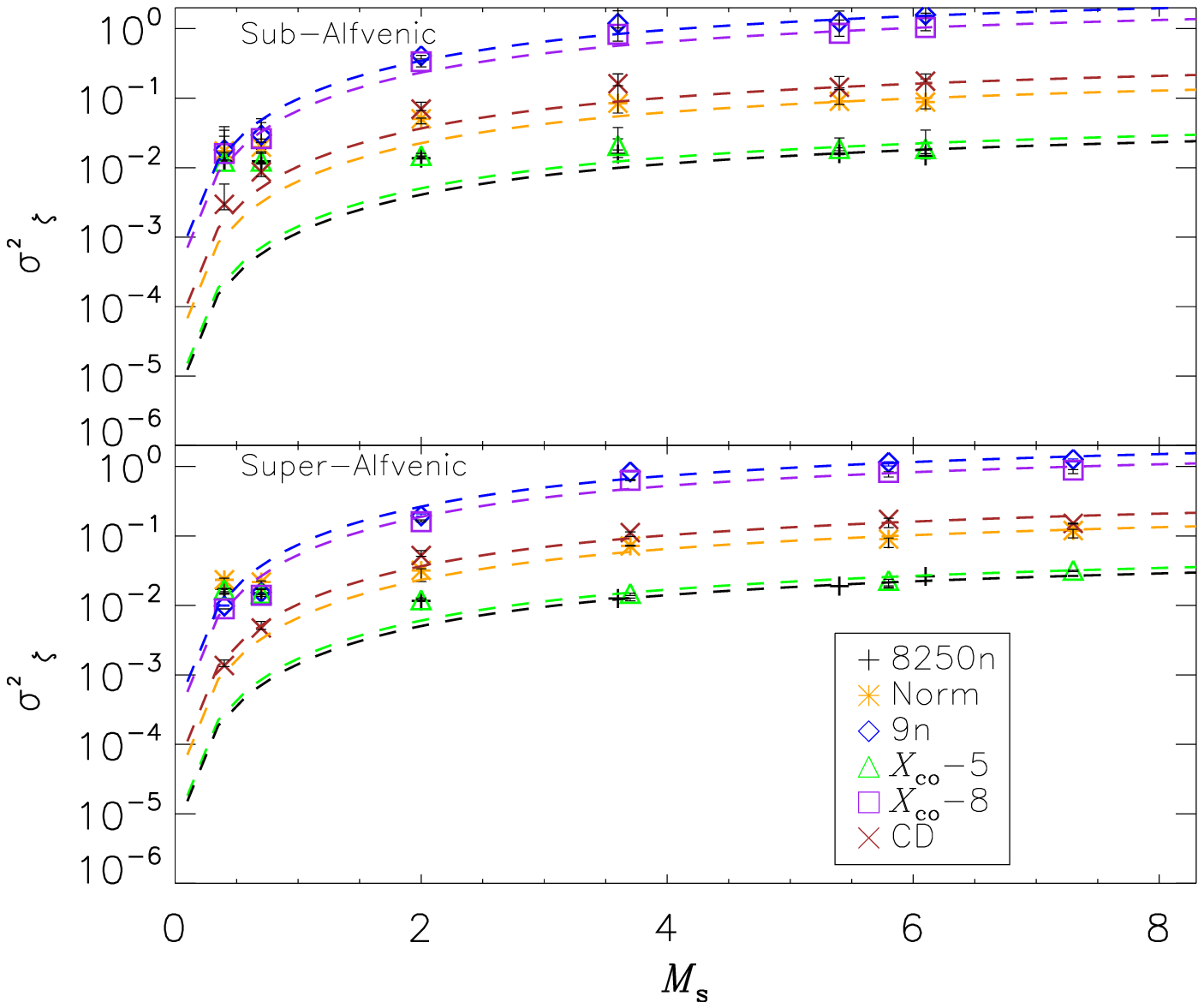}
\caption{Same as Figure 4, but showing ($\sigma^2_{\zeta}$) vs. RMS ${\cal M}_s$ instead of $\sigma^2_{\Sigma/\Sigma_0}$,
for all our simulations with the best fit given by Equation \ref{eq:logvar_cd} (dashed lines).  }
\label{fig:momdn_log}
\end{figure}

\section{M$_s$-$\sigma$ Relation For Column Density and Integrated Intensity Maps}
\label{sec:lin}
In this section we investigate the relationship between the root-mean-squared (RMS) sonic Mach number and the variance of both the linear column density/integrated intensity
and logarithmic column density/integrated intensity. The relationship between the 3D variance and sonic Mach number is given by Equation \ref{eq:linvar}
for density, with parameter $b$ being empirically derived from numerical simulations.  

More recently, Burkhart \& Lazarian (2012) formulated a variant of Equations \ref{eq:linvar} and \ref{eq:logvar} for 2D turbulent maps as:

\begin{equation}
  \sigma_{\Sigma/\Sigma_0}^2=(b^2{\cal M}_s^2+1)^A-1
\label{eq:linvar_cd}
\end{equation}

and 

\begin{equation}
\sigma_{\zeta}^2=A\times ln(1+b^2{\cal M}_s^2)
\label{eq:logvar_cd}
\end{equation}
,where $b=1/3$ (see Federrath et al. 2010) and $A=0.1$ 
were found from a best fit.
However, they did not investigate the effects of radiative transfer that are critical for application to the molecular medium.  In the following subsections, we will
investigate the robustness of Equations \ref{eq:linvar_cd} and \ref{eq:logvar_cd} on our $^{13}$CO integrated intensity maps.

\subsection{Linear Variance vs. Sonic Mach Number}
In Figure \ref{fig:var_lin} we plot the linear variance of ($\sigma^2_{\Sigma/\Sigma_0}$) vs. sonic Mach number 
for all our simulations. The dashed lines show the the best fit given by Equation \ref{eq:linvar_cd} 
including radiative transfer effects and using b=1/3 for solenodially driven simulations  (see Federrath et al. 2010).  Error bars
are created by taking the standard deviation between the variance of integrated intensity maps with line-of-sight  parallel and perpendicular to the mean magnetic field.
The resulting parameter $A$ is given in Table 2. 

Inspection of Figure \ref{fig:var_lin} reveals that
values of $A$, for sonic Mach numbers greater than 2, are clearly dependent on the radiative transfer environment. 
 When we include the effects of radiative transfer, the closest match to the column density
is surprisingly not for optically thin tracers, but rather our norm case with $\tau \approx 1$.
Equation \ref{eq:linvar_cd} provides a good fit to the true column density (red symbols), the mixed optically thin/thick case (yellow line)
and the high optical depth cases (green and black lines) at supersonic Mach numbers.  For subsonic Mach numbers, the  
global radiative coupling for Mach numbers around unity and below distorts the relationship between the sonic Mach number and variance.

\begin{table}
\begin{center}
\caption{Best Fit Parameters for Linear Variance Using Equation \ref{eq:linvar_cd}  }
\begin{tabular}{cccc}
\hline\hline
Model &average $\tau$ & Best fit A & Alfv\'enic Nature\\
\tableline
True Column Density&N/A  & 0.1 & Super-Alfv. \\
8250$n$ &Optically Thick  & 0.01& Super-Alfv.\\
Norm& Mixed &  0.07 & Super-Alfv.  \\
9$n$ & Optically Thin  & 0.7 & Super-Alfv.\\
$\chi-5$ &Optically Thick & 0.02 & Super-Alfv. \\
$\chi-8$&Optically Thin &  0.5 & Super-Alfv. \\
\tableline
True Column Density&N/A  & 0.1 & Sub-Alfv. \\
8250$n$ &Optically Thick  & 0.01 & Sub-Alfv. \\
Norm& Mixed &    0.07  & Sub-Alfv.  \\
9$n$ & Optically Thin  & 0.9  & Sub-Alfv.\\
$\chi-5$ &Optically Thick & 0.01  & Sub-Alfv. \\
$\chi-8$&Optically Thin &  0.7 & Sub-Alfv. \\

\hline\hline
\end{tabular}

\end{center}
\label{tab:twopar}

\end{table}

\subsection{Logarithmic Variance vs. sonic Mach number}
We analyze the variance of the logarithmic integrated intensity maps and their relationship with the sonic Mach number (i.e. we plot $\sigma_\zeta^2$ vs. $M_s$) in Figure \ref{fig:momdn_log}. Error bars
are created by taking the standard deviation between the variance of integrated intensity maps with line-of-sight  parallel and perpendicular to the mean magnetic field.
The dashed lines represent Equation \ref{eq:logvar_cd} when using $ b=1/3$ and the $A$ parameters from the fit of the linear distributions in Table 2. 
The resulting fits are very similar to what was found for the case of the linear variance vs. sonic Mach number.
Equation \ref{eq:logvar_cd} provides an acceptable fit to the true column density (red symbols), the mixed optically thin/thick case (yellow line)
and the high optical depth cases (green and black lines) at supersonic Mach numbers.  For subsonic Mach numbers, again we find that the  
global radiative coupling  distorts the relationship between the sonic Mach number and variance.  
In general, values for $A$ are larger in cases where the optical depth is low and smaller for higher
optical depth.

\section{M$_s$-Skewness and Kurtosis of Column Density and Integrated Intensity Maps}
\label{sec:skew}
The higher order moments skewness and kurtosis (definitions are given in Equations \ref{eq:skew} and \ref{eq:kurt}) have also been shown to have a strong dependency on the
sonic Mach number of both 3D density and 2D column density (Kowal et al. 2007; Burkhart et al. 2009, 2010).

\begin{figure}[tbh]
\centering
\includegraphics[scale=.8]{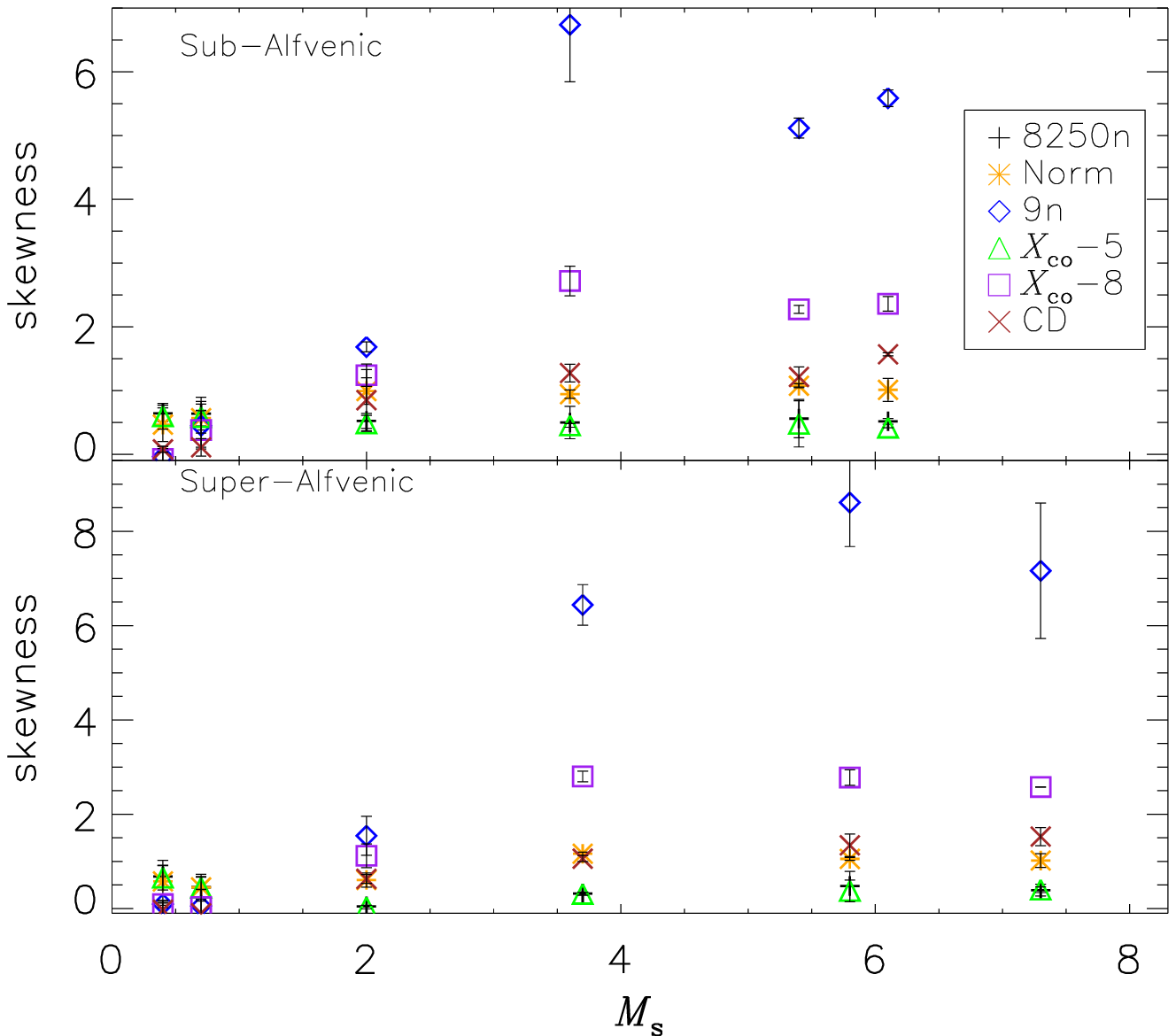}
\caption{Skewness of the  linear distribution ($\Sigma$) vs. RMS sonic Mach number. Sub-Alfv\'enic models are shown top while super-Alfv\'enic models are shown in the bottom. }
\label{fig:momdns}
\end{figure}

\begin{figure}[tbh]
\centering
\includegraphics[scale=.8]{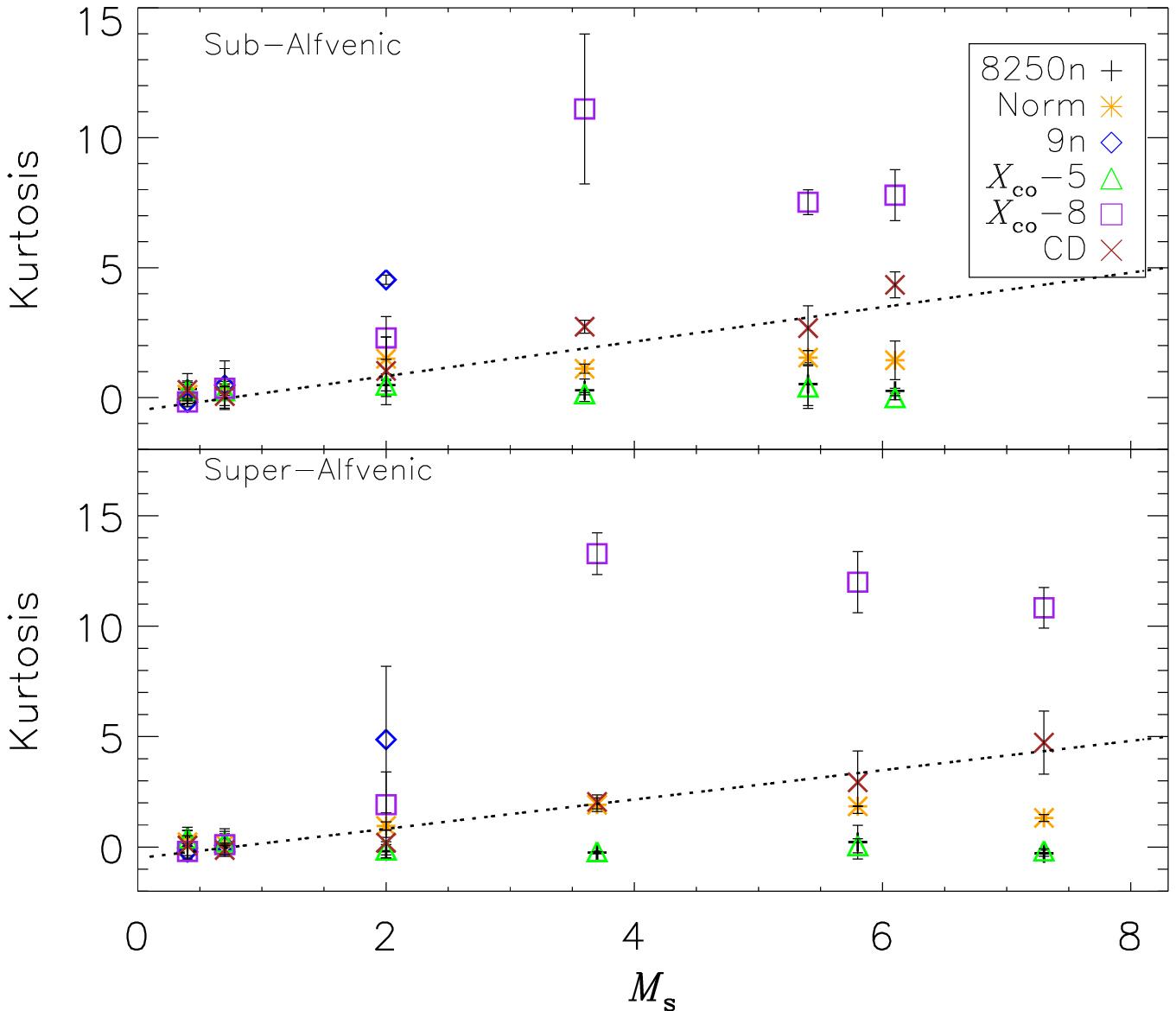}
\caption{Kurtosis of the  linear distribution ($\Sigma$) vs. RMS sonic Mach number. 
Sub-Alfv\'enic models are shown top while super-Alfv\'enic models are shown in the bottom.
The black dotted line is the linear fit used in Burkhart et al. (2010), which follows well the true column density. For ease of viewing, we do not include the kurtosis
of model $9n$  on the plot since their kurtosis values are very high (greater than 60) for the  ${\cal M}_s\approx 4,5,7$ models. }
\label{fig:momdnk}
\end{figure}

We calculate the higher order moments for the linear distribution of the integrated intensity maps and for the true column density
as per Equations \ref{eq:skew} and \ref{eq:kurt}.
We plot the skewness and kurtosis vs. sonic Mach number for the integrated intensity maps  in Figures \ref{fig:momdns} and \ref{fig:momdnk}, respectively. For the kurtosis vs. sonic Mach number,
we over-plot a black dashed line to show the linear fit used in Burkhart et al. 2010). 

For both Figures \ref{fig:momdns} and \ref{fig:momdnk}, in the case of  optically thin media (blue and purple), the higher order moments increase more dramatically (i.e. non-Gaussianity increases)
with increasing sonic Mach number than the true column density.  In particular, the low density case (blue diamonds, model $9n$) departs from the general monotonic trend seen in 
the true column density and other models.  The low density case also has extremely large values of kurtosis (greater than 60) for our highly supersonic models, and we do not show them on the plot in order to 
preserve the visualization for the other models. 
The model that most closely matches the true column density in both skewness and kurtosis is the norm case with $\tau \approx 1$.
As the optical depth increases, the dependency of the sonic Mach number on the skewness and kurtosis becomes less apparent.
In both cases of changing the density and
abundances it is clear that the higher order moments are very sensitive 
to the radiative transfer environment in addition to the sonic Mach number.  

\begin{figure}[tbh]
\centering
\includegraphics[scale=.8]{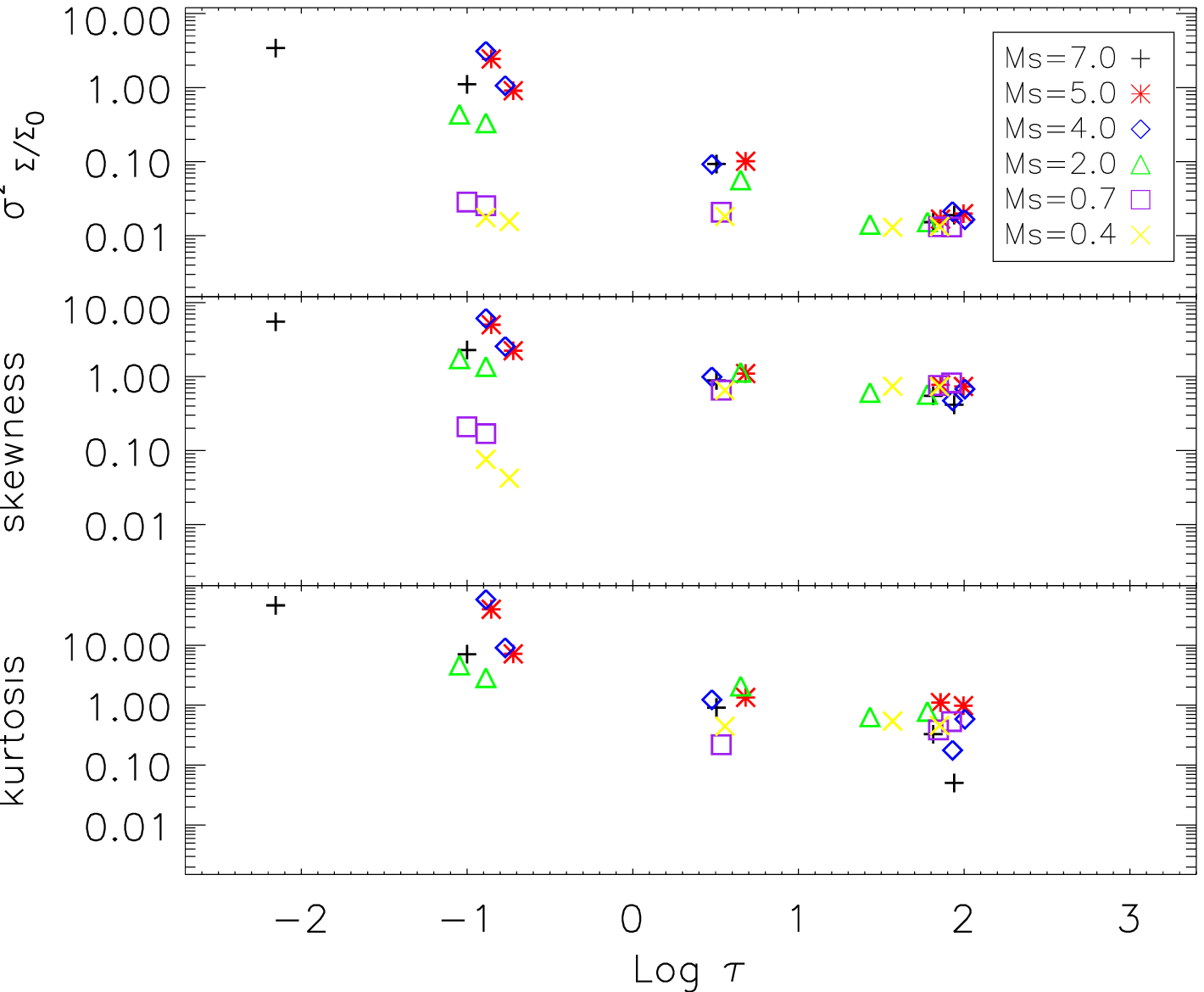}
\caption{Linear moments vs. $\tau$ for sub-Alfv\'enic models on a logarithmic scale.  Panels are ordered from top to bottom as variance, skewness, and kurtosis. Lower values of $\tau$ result in higher values of moments for supersonic turbulence.  For subsonic case, lower $\tau$ results in 
lower values of skewness and kurtosis. In particular, values of kurtosis for subsonic simulations are negative and hence are not plotted here on the log scale.  
The kurtosis values for subsonic simulations with low optical depths are of order -2.     }
\label{fig:tau}
\end{figure}

We chose here to focus on the relationship between the moments of the column density distribution and the sonic Mach number for the CO integrated intensity maps.   In the above analysis, it is clear that an important parameter to take into account when applying PDF-Mach number relations on CO is the optical depth.  In this case we also show how the moments change with varying $\tau$ in Figure \ref{fig:tau} for our sub-Alfv\'enic models.   If the sonic Mach number is known (e.g. via velocity dispersions), then one can potentially estimate the order of magnitude of the optical depth in supersonic clouds. 

\section{TSALLIS STATISTICS}
\label{sec:tsal}
A separate avenue for investigation of PDFs of interstellar turbulence has been to use \textit{incremental PDFs} along with the Tsallis function.
The Tsallis distribution (Tsallis 1988) was originally derived as
a means to extend traditional Boltzmann-Gibbs mechanics to fractal and
multi-fractal systems. 

\begin{figure*}[tbh]
\centering
\includegraphics[scale=.5]{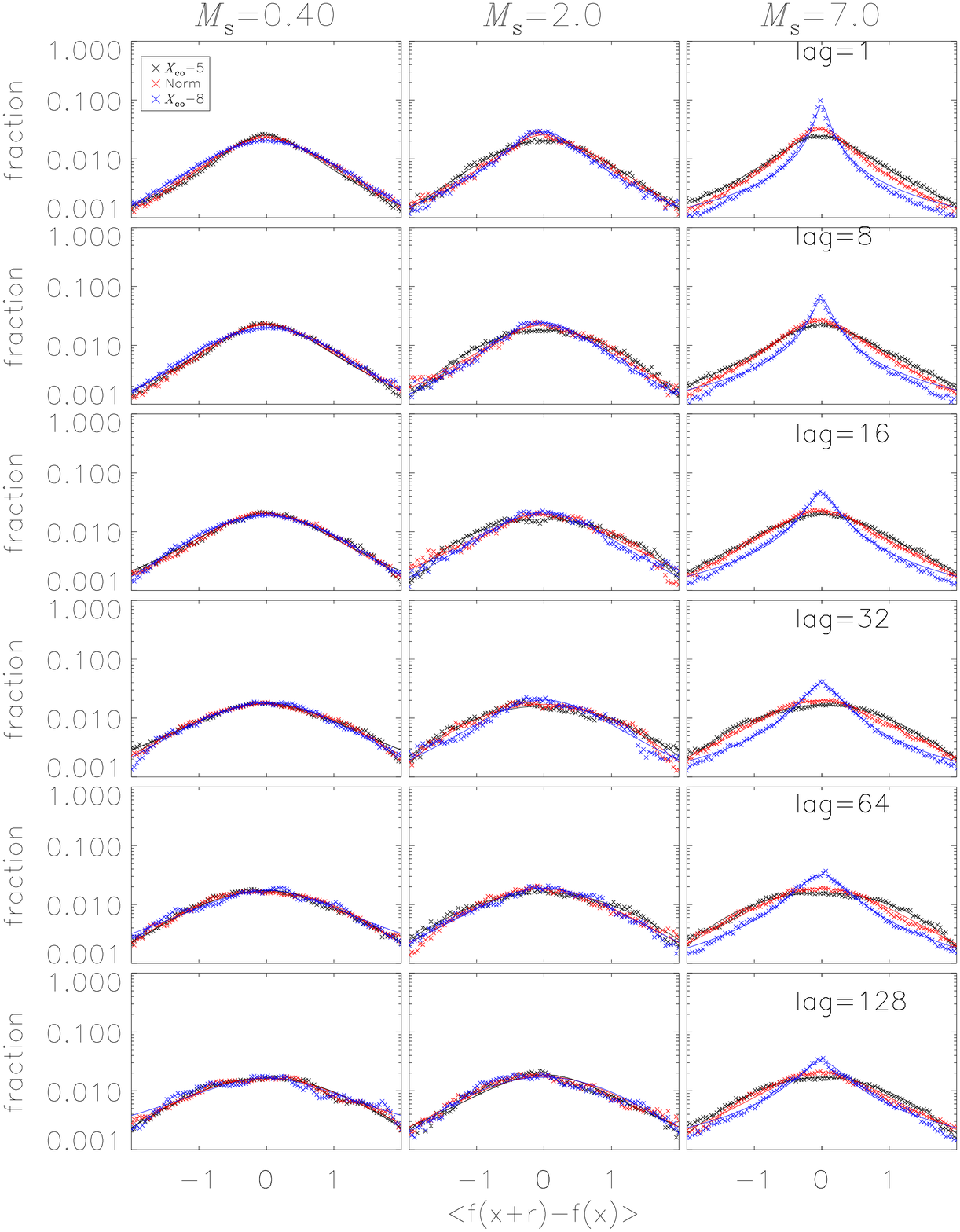}
\caption{Tsallis fit (solid lines) for incremental PDFs of
 super-Alfv\'enic models with varying abundances (as shown in Table \ref{fig:radtab}). The blue line is the optically
 thin case, the red line represents our norm case with $x_{co} = 1.5\times10^{-6}$ and  $n=275 \rm{cm}^{-3}$
 and the black line is the most optically thick case with abundance value $30\times$ larger than the norm. 
The Tsallis function provides good fits  to PDFs of the integrated intensities with varying optical depth. Columns
show different sonic Mach number while rows show different incremental lag values.}
\label{fig:tsalpdf}  
\end{figure*}

The Tsallis function is used to fit incremental PDFs, i.e. PDFs of 
$\Delta f(r)=(f(r)-\langle f(r)\rangle_{\bf{x}})/\sigma_{f}$, where
$\langle$...$ \rangle_{\bf{x}}$
refers to a spatial average. We note that the increments  do not need to be spatial but could
also be temporal increments. 

The Tsallis function of an arbitrary incremental PDF $(\Delta f)$ has the form:

\begin{equation}
\label{(1)}
R_{q}= a \left[1+(q-1) \frac{\Delta f(r)^2}{w^2} \right]^{-1/(q-1)}
\end{equation}
The fit is described by the three parameters $a$, $q$, and $w$. The
$a$ parameter describes the amplitude while $w$ is related to the width
or dispersion of the distribution. Parameter $q$, referred to as the ``non-extensivity
parameter'' or ``entropic index'', describes the sharpness and tail size of the
distribution.

The Tsallis fit parameters are in many ways similar to statistical moments explored in the previous sub-section.
In regards to the Tsallis fitting parameters, the $w$
parameter is similar to the second order moment variance while $q$ is
closely analogous to fourth order moment kurtosis. Unlike higher
order moments, however, the Tsallis fitting parameters are dependent least-squares
fit coefficients and are more sensitive to subtle changes in the PDF. 

 We apply the Tsallis fit to PDFs of increments using the  $^{13}$CO 2-1  integrated intensity maps varying the average density and abundance factor. 
We show an example of the Tsallis  fits for super-Alfv\'enic  $^{13}$CO 2-1 turbulence in Figure \ref{fig:tsalpdf}.
We find that the Tsallis function fits incremental PDFs of integrated intensity maps well, including both regimes of optically
thin and thick. As an example,  we only show the results of varying the abundance in  Figure \ref{fig:tsalpdf} and omit varying the density parameter space.
The blue line is the optically
 thin case, the red line represents our standard case with $x_{co} = 1.5\times10^{-6}$ and  $n=275 \rm{cm}^{-3}$
 and the black line is the most optically thick case with abundance value $30\times$ larger than the norm. 
 Columns show different sonic Mach number while rows show different incremental lag values.
 Solid lines represent the Tsallis fit while the dotted lines are the PDFs. 
 
 We find that the incremental PDFs are less peaked for cases of high lag, high optical depth and low
 sonic Mach number.
Examination of Figure  \ref{fig:tsalpdf} shows that, for the
fit values of the 
width and amplitude of the PDF of the  ${\cal M}_s$=7.0 case shows 
a strong dependency on the optical depth/ radiative transfer environment over all lag values.
Interestingly, we see less of a  dependency on the radiative transfer for the ${\cal M}_s$=2.0 and ${\cal M}_s$=0.4 cases.

\begin{figure}
\centering
\includegraphics[scale=.6]{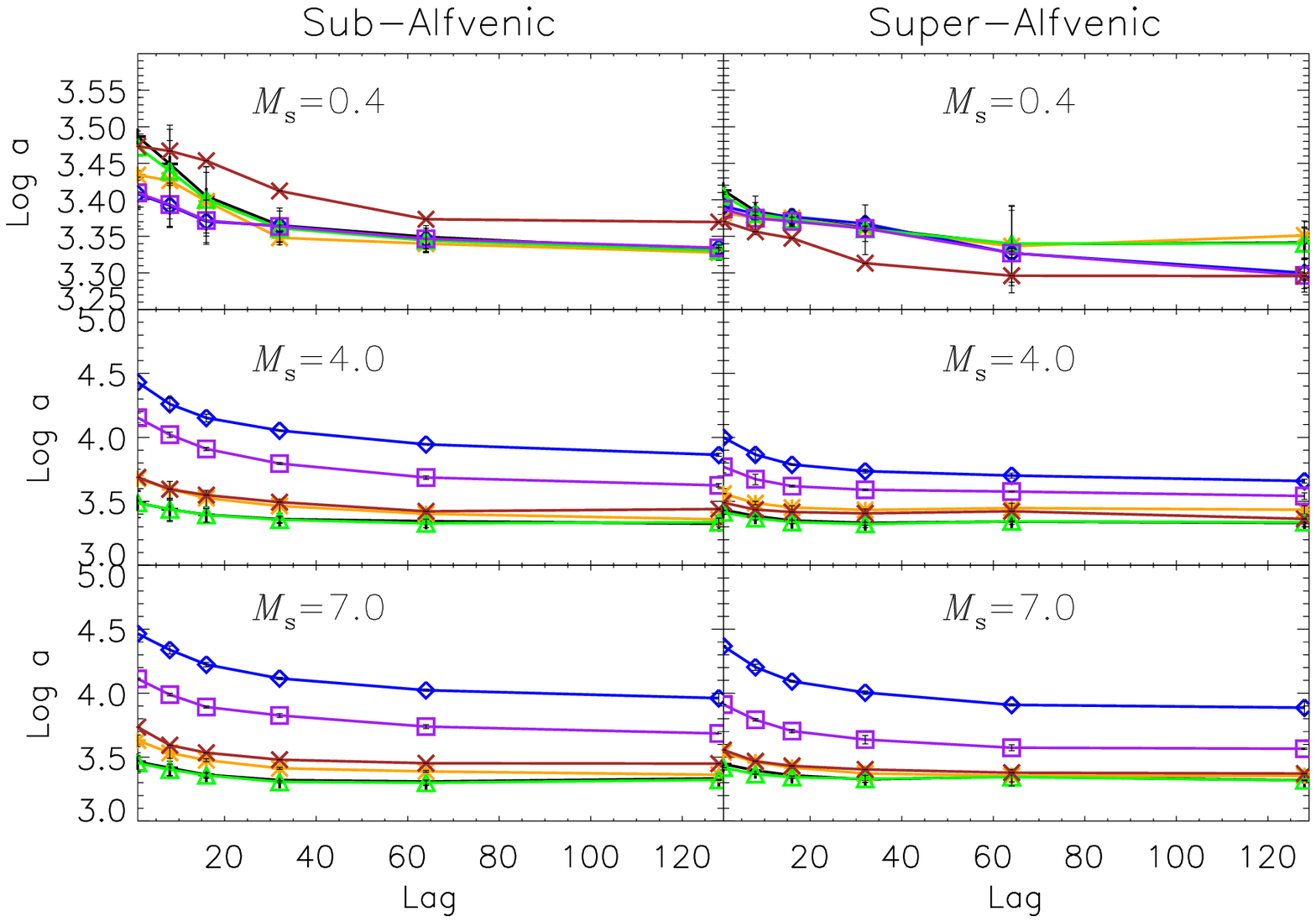}
\includegraphics[scale=.6]{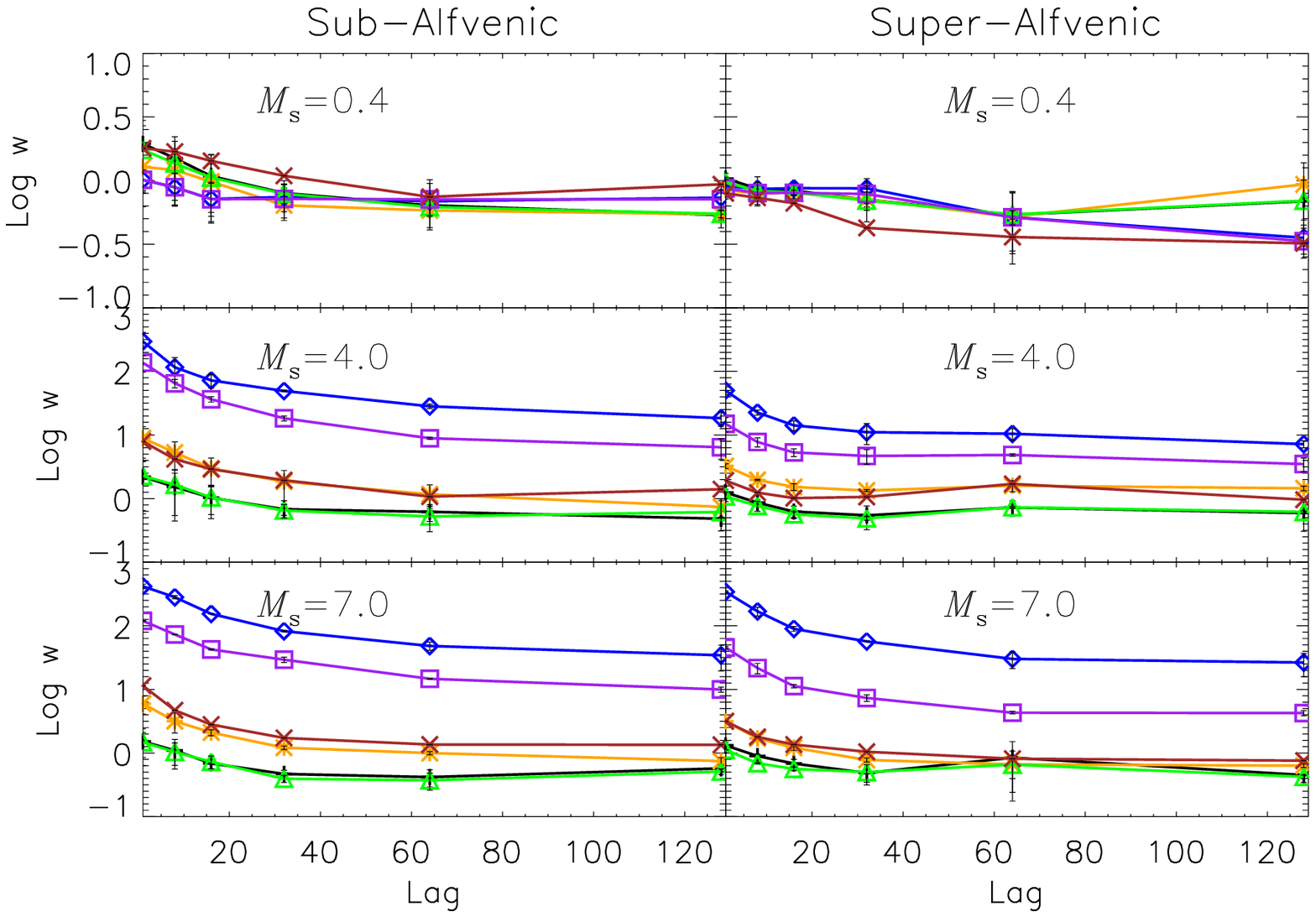}
\caption{Tsallis fit parameters for the amplitude (Log a, top panel) and width (Log w, bottom panel) vs. lag for simulations
with varying average density  and abundance.  Rows show varying increasing sonic Mach number from top to bottom and the two
column are sub- and super-Alfv\'enic left to right, respectively.  The symbols and color scheme are identical to Figure \ref{fig:momdns}. }
\label{fig:tsal_a}  
\end{figure}
We further investigate the dependencies of the incremental PDF width and amplitude on sonic Mach number and optical depth by plotting the log of the width ($w$) and log of the amplitude parameter ($a$) 
vs. spatial lag for cases of varying density and abundance in Figure \ref{fig:tsal_a}.
Rows show varying sonic Mach number while sub- and super-Alfv\'enic  turbulence is shown left and right.
Cases with higher  magnetization (lower Alfv\'enic number) generally have higher values of the width and amplitude.
The symbol and color scheme used is similar to all previous plots used in this paper.

The log of the amplitude parameter is shown in the top panel of Figure \ref{fig:tsal_a}.  
The  ${\cal M}_s$=7.0 case shows the clearest distinction between cases of high and low optical depth over all  lags. 
The amplitude is highest for the cases of high sonic Mach, especially when the medium is optically thin. 
Once the optical depth has increased into the optically thick regime, differences between sonic Mach number become hard to distinguish. 
When the medium is supersonic and the average density is low (blue line, top row) the amplitude of the incremental PDF is much larger than the high abundance case, which
implies the Tsallis function is also sensitive to changes in density/abundance rather than just the optical depth.

The width parameter (in the bottom panel of Figure \ref{fig:tsal_a}) shows similar trends  as seen in the amplitude parameter.  
The  ${\cal M}_s$=7.0 case shows the clearest distinction between the varying radiative transfer parameter space. 
For supersonic turbulence, the widths are highest in the presence of an optically thin medium. 
Once the optical depth has increased into the optically thick regime, differences between sonic Mach number become hard to distinguish, however, there still remain some differences
in Alfv\'enic Mach number.

For both the amplitude and width parameter, the case that most clearly follows the true column density is the norm case with $\tau \approx 1$.  For this case,
differences in both sonic and Alfv\'enic Mach number can be seen.  This makes the  Tsallis function particularly useful for clouds whose optical depth are around unity.  Furthermore, if the sonic Mach number is known and is found to be supersonic, the Tsallis fit can be used to determine the radiative transfer parameter space.

\section{Discussion}
\label{sec:dis}

This study stresses the importance of radiative transfer 
effects in studying the properties of the observable line intensity distribution of the turbulent ISM. 
 They strongly affect the
molecular line maps in turbulent clouds. We restricted the analysis to the
PDFs of the $^{13}$CO J2-1 transition.  
We expect the same qualitative behavior with higher transitions rather following the reduced density case and 
$^{12}$CO following the high abundance case. 
While we focus our
study on carbon monoxide, our conclusions may well be applicable to other species, e.g HI. 

This study is complementary to a study done by Shetty et al. (2011a,b), which found that the X factor depends strongly on the average density and CO abundances.  As was observed with the PDFs in this work, 
high values of volume density and/or CO abundance can cause the CO line to saturate, which creates integrated intensity maps that miss important structure in the true column density map.
In the case of optically thin CO and low density, the lowest CO density/abundance regions will not collisionally excite, and again we miss out on important
structure in the true column density map.  This harkens back to a study by Goodman et al. (2009b), who showed that one of the biggest downfalls of carbon monoxide tracers is its 
limited dynamical range.  Goodman et al. (2009b) pointed out that dust, while more limited in terms of resolution, has a much larger dynamical range with which to trace out the true column 
density map. However, it cannot trace any velocity information through line profiles.

In this study, we find that CO is able to trace the PDF statistics of the true column density
best for case with $\tau \approx 1$ using average interstellar values for the density ($n=275 cm^{-3}$) and abundances ($x_{co} \approx 10^{-6}$, Burgh et al. 2007), as represented in our $\tau$ norm case.  

It may seem counter-intuitive that the best tracer of the true column density PDF is the one with
$\tau \approx$ unity, since many past studies have always opted for the most optically thin observations available in order to avoid saturation effects.
However, it seems the key issue is the limited dynamical range of CO.  
Figure \ref{fig:supercods} shows that the dynamical range
is higher in the cases of the optically thick cases (due to the fact that these cases are dominated by dispersion).  However, 
the case with optical depth on order unity (central panels of Figure \ref{fig:supercods} shows that this case still traces most of the true column density
structure while having the largest dynamic range. In this case, getting the true column density can no longer be considered only in terms of obtaining the 
correct  X factor (which has its own dependencies on abundance, temperature, velocity range and volume density, see Shetty et al. 2011a,b), but 
also on maximizing the dynamic range.  This is critical for estimating the statistics of the underlying gas column density.

The PDFs of CO maps will be most informative to use when the optical depth of the tracer is known,
as this will provide more certainty on their relationship with the parameters of turbulence.  Conversely, if the sonic Mach number
is known (and is supersonic), then one can get an order of magnitude estimate of the optical depth from the use of the statistics discussed in this work.
Each tool has its own strengths and weaknesses.  The Tsallis statistic is in general, more sensitive to the Alfv\'enic Mach number (especially in the case
of strongly supersonic turbulence) than the moments and more sensitive to the changing radiative transfer parameter space.  Ultimately, a use of multiple well-founded techniques,
coupled with numerical simulations, will provide the most accurate parameters of the turbulence
that exists in the observations.
While PDFs were explored in this paper, 
they are not the only useful tool for studies of turbulence (see the Introduction).
Therefore, in a subsequent paper we will extend the analysis to 
the impact of the radiative transfer on the spatial structure and spatial correlations.

Another question raised by this study is: 
could the PDF methods discussed in this work be used to find the observational sonic and Alfv\'enic Mach numbers, given that the optical depth is known?
To answer this, we generate a new synthetic observation (with numerical resolution of $512^3$) as our test case.  
It has parameters ${\cal M}_s$=2.8,  ${\cal M}_A$=2.0,  $x_{co} \approx 5\times10^{-7}$ and $n=400 cm^{-3}$ which are different from all the other
models listed in Table \ref{fig:radtab} with which we used to draw our conclusions.  
However, if these parameters are now unknown, and all the information we have is the Figures of this paper, the integrated intensity map, and the 
fact that this case has an optical depth $\tau=1.3$, what can we learn?

First, we can easily compute the statistical moments of the linear and logarithmic distributions.  We find that the linear
variance, skewness and kurtosis are $\sigma^2_{\Sigma/\Sigma_0}=$0.11, $\gamma=1.01$, and $\beta=1.38$.  The logarithmic variance is
$\sigma^2_{\zeta}=0.104$.  Using the  calculated values for the variance along with Equations \ref{eq:linvar_cd} and Equations \ref{eq:logvar_cd}
with A=0.07 as given for the norm case (which has a slightly higher optical depth but is the closes in terms of the optical depth parameter space)
in Table 2 provides a value of ${\cal M}_s$=3.0.
Using the linear fit for kurtosis given in  Burkhart et al. 2010 and shown in Figure \ref{fig:momdnk} we find  ${\cal M}_s$=2.7.  As this is very close to the expected sonic Mach number, this  again shows that the moments of cases with $\tau=1$ follow closely the statistics of the true column density.
The skewness of the orange lines in Figure \ref{fig:momdns}, also shows values consistent with the ranges given by skewness.
For the estimation of the Alfv\'enic Mach number, we found that only incremental PDFs (i.e. the Tsallis fit) are substantially sensitive to ${\cal M}_A$, whereas the
fully PDFs of the integrated intensity maps had primary sensitivities to the radiative transfer parameter space and sonic Mach number.
We feel that multiple techniques are required in order to obtain both sonic and Alfv\'enic Mach numbers.  This is especially true
in the case of the Alfv\'enic
Mach number, as many statistics have a primary sensitivity to the compressibility with the magnetic field contributing as a secondary influence.
For example, future works should test how the velocity centroid anisotropy  (see Esquivel \& Lazarian 2011) is affected by radiative transfer.

This test case demonstrates that it is possible to determine parameters of turbulence with the use of statistics on $^{13}$CO data
given information about the optical depth.
In our case, we also had information about the driving of the turbulence in order to use parameter $b=1/3$. 
Studies with different driving (i.e. b ranging from 1/3 to unity) should be considered in the future to assess the sensitivity of the present results on the driving mode of the turbulence.
We stress that the reverse case of knowing
the sonic Mach number, which can be provided by velocity dispersions (e.g. Kainulainen \& Tan 2012), one can obtain information on the optical depth via use of PDFs. Ultimately,
the more statistics that are used together, the more accurate the determination of the parameters of turbulence and radiative transfer environment will be. 
On this point, we note that PDFs represent only a small subset of statistics one can use to study turbulence with.

\section{Conclusions and Summary}
\label{sec:con}
We investigate the PDF moments and Tsallis statistics of  MHD turbulence including radiative transfer effects of the $^{13}$CO 2-1 transition in the
context of their sensitivity to the sonic and Alfv\'enic Mach numbers and the radiative transfer environment (i.e. varying density and abundance values).  
We find that PDF descriptors applied to the integrated
intensity maps generally show a large dependency on the radiative transfer parameter space in addition to the Mach numbers.   
In particular, we find that:
\begin{itemize}
\item PDFs of $^{13}$CO integrated intensity maps are generally log-normal, although deviations are present. 
\item The linear and logarithmic variance-${\cal M}_s$ relationship is observed in integrated intensity maps with optical depths an order of magnitude one or less.
For cases with high CO abundances and densities and therefore optically thick lines, the variance looses sensitivity to the sonic Mach number.  
We recommend using 
the 
 logarithmic variance for estimation of the sonic Mach number, as it is less sensitive to scaling (i.e. choice of X factor) than the linear variance.
\item The higher order moments (skewness and kurtosis) of the integrated intensity distribution depend on the radiative transfer environment
and sonic Mach number. 
For cases where the CO abundance and density values are too low to excite the full dynamic range of the gas,
the skewness and kurtosis increase with increasing sonic Mach number more rapidly than the true column density.
For very optically thick cases, the higher order moments lose their sensitivity to the sonic Mach number. 
\item The Tsallis fit parameters also show sensitivity to the radiative transfer environment however, 
dependencies on both sonic and Alfvenic Mach number exist in cases with $\tau \approx$ an order of magnitude one or less. 
The Tsallis parameters and fit 
follow 
the true column density best for the $\tau \approx 1$ case.  
\end{itemize}

\acknowledgments
B.B. acknowledges support from the NSF Graduate Research Fellowship and the NASA Wisconsin Space Grant
Institution. A.L. thanks both NSF AST 0808118 and the Center for Magnetic Self-Organization in Astrophysical and Laboratory Plasmas for financial support.
All authors acknowledge support through grant SFB956/DFG and by the Deutsche Forschungsgemeinschaft, DFG, project number Os 177/2-1.
This work was completed  during the stay of A.L. as
Alexander-von-Humboldt-Preistr\"ager at the Ruhr-University Bochum and the University of Cologne.

\section{Appendix}

\begin{figure}
\centering
\includegraphics[scale=.6]{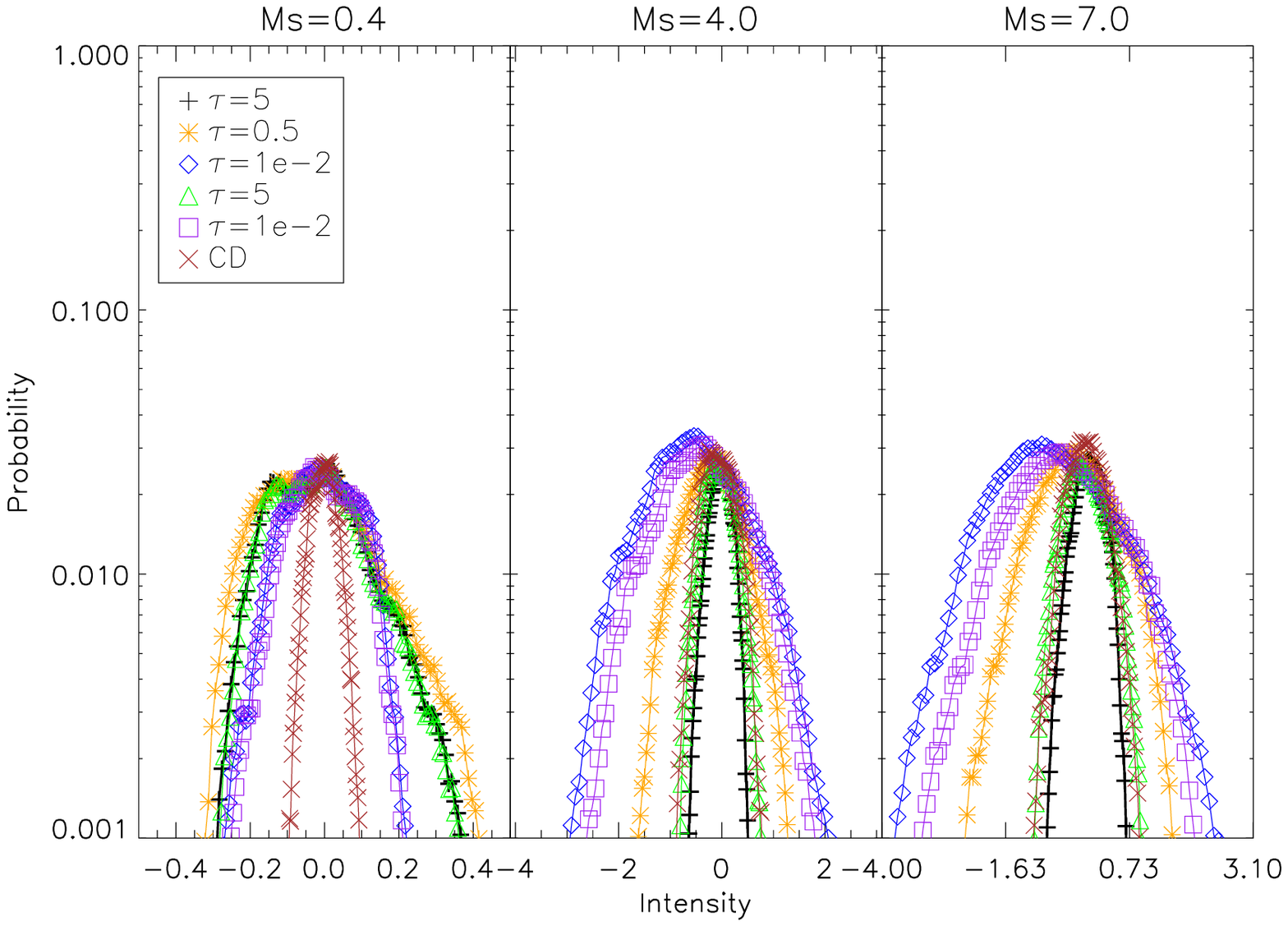}
\caption{PDFs of $\zeta$=ln($\Sigma/\Sigma_0$) for simulations where we scale the velocity field rather than the sound speed in order to vary the sonic Mach number. The far left panel has ${\cal M}_s \approx 0.4$ the central panel has ${\cal M}_s \approx 4.0$, and the far right panel has ${\cal M}_s \approx7.0$.  All models have ${\cal M}_A \approx 2.0$. Results obtained are similar to those in Figure~\ref{fig:lognorm_pdf} in that the synthetic $^{13}$CO integrated intensity maps in with optical depth on order of unity have PDFs that are most similar to the column density PDF.  Optically thin PDFs show larger variance than the true column density.  }
\label{fig:tautest}  
\end{figure}

As mentioned in Section \ref{sec:Sims} the original simulations set the sound speed by
varying the pressure and holding the RMS velocity constant. However, now we set a constant temperature of 10K for all the simulations.
In this case, it makes sense to hold the sound speed constant and vary the RMS velocity in order to set the isothermal sonic Mach number. Here we rescale
the velocity field of all the simulations in order to hold the sound speed constant for a given sonic Mach number.  Changing the velocity field changes the overall radiative transfer calculation, however we find that the resulting PDFs and their dependencies on the sonic Mach number and optical depth are not affected.   This is because the PDFs of the integrated intensity maps are not sensitive to velocity information, and are primarily sensitive to the optical depth and sonic Mach number. We demonstrate this in Figure \ref{fig:tautest}, which is similar to Figure~\ref{fig:lognorm_pdf} except here we use simulations that vary the sonic Mach number via scaling the velocity field rather than the sound speed.   We use the same radiative transfer parameter space (i.e. the same values of density and abundance) that was used in Figure~\ref{fig:lognorm_pdf} , however, because we alter the velocity field, the radiative transfer problem changes as well.  Nevertheless, the behavior of the PDFs as a function of sonic Mach number and optical depth is unchanged compared with Figure~\ref{fig:lognorm_pdf}.  In both cases, the synthetic observations with optical depth on order unity display the most similarity to the column density PDF, while the cases with very low optical depth have PDF widths that are wider than the column density.  

\end{document}